\documentclass[superscriptaddress,showkeys,twocolumn,longbibliography,10pt]{revtex4-1}
\usepackage{graphicx}
\usepackage{amsmath}
\usepackage{mathtools}
\usepackage{bbm}
\usepackage{color}
\usepackage{datetime}
\usepackage{fancyhdr}
\usepackage{mathrsfs}

\makeatletter
\let\save@mathaccent\mathaccent
\newcommand*\if@single[3]{%
  \setbox0\hbox{${\mathaccent"0362{#1}}^H$}%
  \setbox2\hbox{${\mathaccent"0362{\kern0pt#1}}^H$}%
  \ifdim\ht0=\ht2 #3\else #2\fi
  }
\newcommand*\rel@kern[1]{\kern#1\dimexpr\macc@kerna}
\newcommand*\widebar[1]{\@ifnextchar^{{\wide@bar{#1}{0}}}{\wide@bar{#1}{1}}}
\newcommand*\wide@bar[2]{\if@single{#1}{\wide@bar@{#1}{#2}{1}}{\wide@bar@{#1}{#2}{2}}}
\newcommand*\wide@bar@[3]{%
  \begingroup
  \def\mathaccent##1##2{%
    \let\mathaccent\save@mathaccent
    \if#32 \let\macc@nucleus\first@char \fi
    \setbox\z@\hbox{$\macc@style{\macc@nucleus}_{}$}%
    \setbox\tw@\hbox{$\macc@style{\macc@nucleus}{}_{}$}%
    \dimen@\wd\tw@
    \advance\dimen@-\wd\z@
    \divide\dimen@ 3
    \@tempdima\wd\tw@
    \advance\@tempdima-\scriptspace
    \divide\@tempdima 10
    \advance\dimen@-\@tempdima
    \ifdim\dimen@>\z@ \dimen@0pt\fi
    \rel@kern{0.6}\kern-\dimen@
    \if#31
      \overline{\rel@kern{-0.6}\kern\dimen@\macc@nucleus\rel@kern{0.4}\kern\dimen@}%
      \advance\dimen@0.4\dimexpr\macc@kerna
      \let\final@kern#2%
      \ifdim\dimen@<\z@ \let\final@kern1\fi
      \if\final@kern1 \kern-\dimen@\fi
    \else
      \overline{\rel@kern{-0.6}\kern\dimen@#1}%
    \fi
  }%
  \macc@depth\@ne
  \let\math@bgroup\@empty \let\math@egroup\macc@set@skewchar
  \mathsurround\z@ \frozen@everymath{\mathgroup\macc@group\relax}%
  \macc@set@skewchar\relax
  \let\mathaccentV\macc@nested@a
  \if#31
    \macc@nested@a\relax111{#1}%
  \else
    \def\gobble@till@marker##1\endmarker{}%
    \futurelet\first@char\gobble@till@marker#1\endmarker
    \ifcat\noexpand\first@char A\else
      \def\first@char{}%
    \fi
    \macc@nested@a\relax111{\first@char}%
  \fi
  \endgroup
}
\makeatother

\makeatletter
\renewcommand{\p@subsection}{}
\renewcommand{\p@subsubsection}{}
\makeatother

\newcommand{\gone}[1]{}
\newcommand{\eqgone}[1]{}
\renewcommand{\eqgone}[1]{#1}
\newcommand{\figsgone}[1]{}
\renewcommand{\figsgone}[1]{#1}
\newcommand{\clearpf}{\ \clearpage}
\renewcommand{\clearpf}{}
\newcommand{\ttume}{\affiliation{Department of Mechanical Engineering,
    Texas Tech University, Lubbock, Texas, U.S.A.}}
\newcommand{\ttumephys}{\affiliation{Departments of Mechanical Engineering
    and Physics, Texas Tech University, Lubbock, Texas, U.S.A.}}
\newcommand{\ttuce}{\affiliation{Department of Chemical Engineering,
    Texas Tech University, Lubbock, Texas, U.S.A.}}
\newcommand{\torsional}{\tau}
\newcommand{\curvatural}{c}
\newcommand{\roll}{{\textrm{roll}}}
\newcommand{\turn}{{\textrm{turn}}}
\newcommand{\attr}{t}
\newcommand{\ctrl}{c}
\newcommand{\burrowing}{b}
\newcommand{\swimming}{s}
\newcommand{\low}{L}
\newcommand{\high}{H}
\newcommand{\bcdot}{\boldsymbol{\cdot}}
\newcommand{\variance}{\sigma}
\newcommand{\frequency}{f}
\newcommand{\esm}{ESM}
\newcommand{\ESM}{Electronic Supplementary Material}

\newcommand{\threed}{3D}
\newcommand{\twod}{2D}
\newcommand{\celegans}{{\it C. elegans}}
\newcommand{\Celegans}{{\it Caenorhabditis elegans}}
\newcommand{\Cshaped}{\textit{C}-shaped}
\newcommand{\Sshaped}{\textit{W}-shaped}

\newlength{\textwidthF}
\newcommand{\figsize}{0.45\textwidth}
\newcommand{\figsizeGraph}{0.46\textwidthF}

\newcommand{\curvature}{\kappa}
\newcommand{\torsion}{\tau}

\newcommand{\wormcurvature}{\curvature_w}
\newcommand{\wormlength}{L}
\newcommand{\wormcoord}{s}
\newcommand{\curvecoord}{\wormcoord^\prime}
\newcommand{\chronos}{t}
\newcommand{\wavespeed}{v_s}
\newcommand{\amplitude}{A}
\newcommand{\wavevector}{q}

\newcommand{\phase}{\phi}

\newcommand{\utanvec}{{\hat{\boldsymbol{t}}}}
\newcommand{\unorvec}{{\hat{\boldsymbol{n}}}}
\newcommand{\ubinvec}{{\hat{\boldsymbol{b}}}}

\newcommand{\turnangleBelly}{\theta^\roll}
\newcommand{\turnvelBelly}{\omega^\roll}
\newcommand{\turnanglePlanar}{\theta^\turn}
\newcommand{\offp}{^\torsional}
\newcommand{\onp}{^\curvatural}

\newcommand{\Ci}{\textit{CI}}
\newcommand{\concentration}{C}
\newcommand{\response}{M}
\newcommand{\dtime}{\chronos'}
\newcommand{\csignal}{Q}
\newcommand{\decay}{\epsilon}
\newcommand{\turnrate}{r_s}
\newcommand{\turnsensitivity}{\alpha_s}
\newcommand{\corkrate}{r_\torsional}

\newcommand{\turnratelow}{r_s^{\low}}
\newcommand{\turnratehigh}{r_s^{\high}}
\newcommand{\turnrateconst}{r_s^0}
\newcommand{\corkrateconst}{r_\torsional^0}
\newcommand{\noworms}{n}
\newcommand{\nowormsA}{\noworms_\attr}
\newcommand{\nowormsC}{\noworms_\ctrl}

\newcommand{\gaussWidth}{\sigma}
\newcommand{\dishWidth}{W}

\begin{document}
\title
    {  
   Navigation of \textbf{ \emph{C.  elegans} } in
   three-dimensional media: roll maneuvers and planar turns
}
\author{Alejandro Bilbao} \ttume
\author{Amar K. Patel} \ttume
\author{Mizanur Rahman} \ttuce
\author{Siva A.\ Vanapalli} \ttuce
\author{Jerzy Blawzdziewicz} \ttumephys

\begin{abstract}
Free-living nematode \Celegans\ is a powerful genetic model, essential
for investigations ranging from behavior to neuroscience to aging, and
locomotion is a key observable used in these studies.  However,
despite the fact that in its natural environment \celegans\ moves in
three-dimensional (\threed) complex media (decomposing organic matter
and water), quantitative investigations of its locomotion have been
limited to two-dimensional (\twod) motion.  Based on our recent
quantitative analysis of \twod\ turning maneuvers [Phys.\ Fluids 25,
  081902 (2013)] we follow with the first quantitative description of
how \celegans\ moves in \threed\ environments.  We show that by
superposing body torsion and \twod\ undulations, a burrowing or
swimming nematode can rotate the undulation plane. A combination of
these roll maneuvers and 2D turns associated with variation of
undulation-wave parameters allows the nematode to explore
\threed\ space.  We apply our model to analyze \threed\ chemotaxis of
nematodes burrowing in a gel and swimming in water; we conclude that
the nematode can achieve efficient chemotaxis in different
environments without adjusting its sensory-motor response to chemical
signals.  Implications of our findings for
understanding of \threed\ neuromuscular control of nematode body are
discussed.

\keywords{\Celegans, locomotion, hydrodynamic bead models, chemotaxis,
  neuromuscular control, 3D environment}
\end{abstract}


   \maketitle
\clearpf
\section{Introduction}

Investigations of locomotion of a free-living nematode \celegans\ (one
of the most important model organisms) can reveal a wealth of
information on physiology, muscle biology, and neural control of
movement and behavior.  Efficient locomotion depends on muscular
strength
\cite{%
  Ghanbari-Nock-Johari-Blaikie-Chen-Wang:2012,%
  Johari-Nock-Alkaisi-Wang:2013,%
  Etheridge-Rahman-Gaffney-Shaw-Shephard-Magudia-Solomon-Milne-Blawzdziewicz-Constantin-Teodosiu-Greenhaff-Vanapalli-Szewczyk:2015},
neuromuscular coordination
\cite{%
  Wen-Po-Hulme-Chen-Liu-Kwok-Gershow-Leifer-Butler-Fang-Yen-Kawano-Schafer-Whitesides-Wyart-Chklovskii-Zhen-Samuel:2012,%
  Stirman-Crane-Husson-Wabnig-Schultheis-Gottschalk-Lu:2011},
and information processing
\cite{%
  Gray-Hill-Bargmann:2005,%
  Kocabas-Shen-Guo-Ramanathan:2012%
}.                        
Locomotory readouts are thus extensively used to describe impacts of 
genetic mutations
\cite{%
   Sznitman-Purohit-Krajacic-Lamitina-Arratia:2010,%
   Brown-Yemini-Grundy-Jucikas-Schafer:2013,%
   Restif-Ibanez_Ventoso-Vora-Guo-Metaxas-Driscoll:2014%
}
and evaluate pharmacological effects
\cite{Artal_Sanz-deJong-Tavernarakis:2006,%
  Omura-Clark-Samuel-Horvitz:2012}.

In its natural environment (soil rich in decaying organic matter),
\celegans\ interacts with a complex three-dimensional (\threed)
surrounding medium that includes rainwater and soft organic materials.
Most studies of nematode locomotion, however, have focused on
two-dimensional (\twod) motion
\cite{Cohen-Boyle:2010,%
Fang_Yen-Wyart-Xie-Kawai-Kodger-Chen-Wen-Samuel:2010,%
Sznitman-Shen-Purohit-Arratia:2010,%
Padmanabhan-Khan-Solomon-Armstrong-Rumbaugh-Vanapalli-Blawzdziewicz:2012,%
Stephens-Johnson-Kerner-Bialek-Ryu:2008},
and little effort has been made to characterize worm motility in
\threed\ environments that would mimic \celegans\  natural habitat.

Only recently was \threed\ burrowing motion of \celegans\ in soft
substrates quantitatively imaged
\cite{Kwon-Pyo-Lee-Je:2013}
and feasibility  of using \threed\  motion to assess genetic mutations
\cite{Kwon-Hwang-You-Lee-Je:2015}
and to identify muscular dystrophy
\cite{Beron-Vidal-Gadea-Cohn-Parikh-Hwang-Pierce-Shimomura:2015}
experimentally demonstrated.  Such \threed\ methods of scoring
\celegans\ locomotory behavior are promising; however, unleashing the
full potential of such techniques requires complementary modeling
efforts to elucidate the biomechanics and neuromuscular control of 3D
locomotion.

Very few quantitative analyses of \threed\ locomotion
of \celegans\ have been reported
\cite{Deng-Xu:2014,%
  Delmotte-Climent-Plouraboue:2015};
no investigations present a  realistic mathematical
description of nematode motion.  Our study aims to fill this knowledge
gap by providing quantitative understanding of the geometry and
mechanics of \threed\ maneuvers of \celegans\ burrowing in gel-like
materials and swimming in Newtonian fluids.

Our modeling approach generalizes the recently proposed piecewise
harmonic curvature (PHC) model of \twod\ nematode gait
\cite{Padmanabhan-Khan-Solomon-Armstrong-Rumbaugh-Vanapalli-Blawzdziewicz:2012}.
The PHC model has been shown to give an accurate parameterization
of body postures (defined by the configuration of
the body centerline
\cite{%
  Stephens-Johnson-Kerner-Bialek-Ryu:2008,%
  Sznitman-Shen-Sznitman-Arratia:2010,%
  Husson-Costa-Schmitt-Gottschalk:2012,%
  Berman-Kenneth-Sznitman-Leshansky:2013})
and of entire trails of nematodes crawling on agar without sidewise
slip; it also yields a convenient quantification of time-lapse
sequences of body postures in swimming.

In the present paper we expand the PHC gait model by including curve
torsion in the description of the nematode postures.  We show that
combining the PHC model for the line curvature with a piecewise
constant torsion (PCT) assumption, we can generate three-dimensional
trajectories that mimic paths of burrowing nematodes.  Using, in
addition, our hydrodynamic models of swimming dynamics
\cite{Bilbao-Wajnryb-Vanapalli-Blawzdziewicz:2013}, we can describe 3D
motion of \celegans\ in bulk fluids. In particular, our gait model
reproduces the rapid undulation-plane reorientation observed
in our experiments.  

The explicit PHC representation of the nematode gait has been shown to
be a useful tool for quantitative investigations of swimming
efficiency and \twod\ turning maneuvers
\cite{Bilbao-Wajnryb-Vanapalli-Blawzdziewicz:2013}.  The combined
PHC/PCT gait model will help analyze diverse aspects of
\threed\ undulatory locomotion of \celegans\ in complex media,
including chemotaxis and mechanisms of neuro-muscular actuation of
body movements.  To illustrate the former application we present an
analysis of the efficiency of \threed\ chemotaxis of burrowing and
swimming \celegans.

Our paper is organized as follows.  In Sec.\ \ref{crawling section} we
develop geometrical gait models for \celegans\ moving in \twod\ and
\threed.  We also assess the effectiveness of reorientation maneuvers
of animals crawling on a flat \twod\ substrate and burrowing in soft
3D matter.  In Sec.\ \ref{maneuverability in fluids} we present our
hydrodynamic analysis of reorientation maneuvers of
\celegans\ swimming in a viscous fluid.  The geometrical and
hydrodynamic locomotion descriptions developed in Secs.\ \ref{crawling
  section} and \ref{maneuverability in fluids} are combined in
Sec.\ \ref{chemotaxis in 3d} with simple models of chemosensation and
motion control to describe chemotaxis of burrowing and swimming
\celegans\ in \threed.  Implications of our findings for the emerging
area of \threed\ undulatory locomotion are discussed in
Sec.\ \ref{Summary and discussion}.

\figsgone{
\clearpf

\begin{figure}
\includegraphics[width=0.34\textwidth]{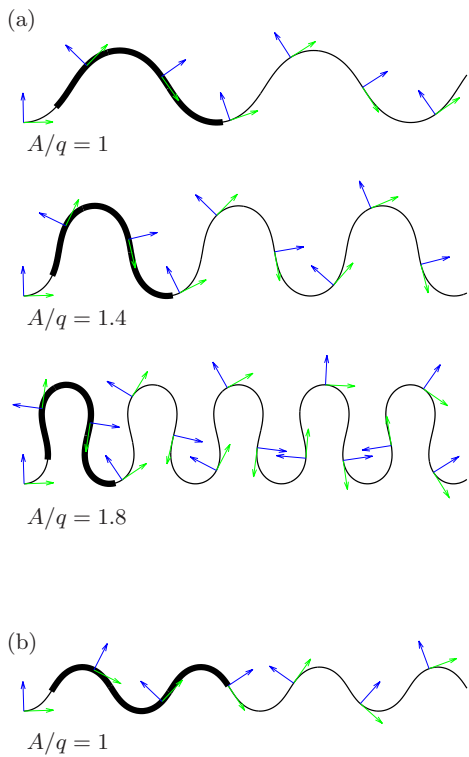}
\caption
{ ({\it Colour online}) Family of curves reproduced from harmonic
curvature \eqref{single mode curvature} for normalized amplitude
$\amplitude/\wavevector$ as labeled.  The green and blue arrows
are the unit tangent and normal vectors, respectively.  The body
of a nematode moving along the curves is represented by the thick
line segments of length $\wormlength$. Two normalized wavevectors
are shown (a) $\wavevector\wormlength=5.5$ (typical
swimming gait) and (b) $\wavevector\wormlength=9$ (typical
crawling gait).  Depending on the wavevector, the
normalized amplitude $\amplitude/\wavevector=1$ may correspond to
(a) \Cshaped\ (top) and (b) \Sshaped\ body postures;
$\amplitude/\wavevector=1.8$ produces an $\Omega$-shaped body
posture.
\label{curves with no torsion}
}
\end{figure}

\begin{figure}
\includegraphics[width=\figsize]{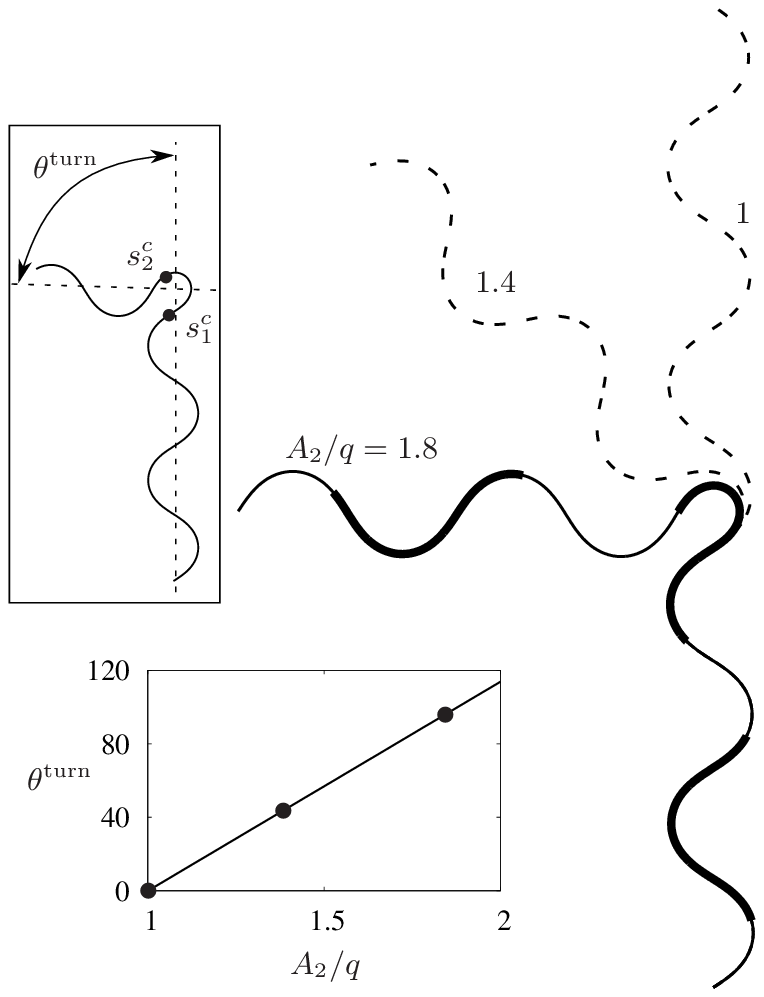}
\caption{ A crawling nematode performing three-mode planar turn
with the default-mode amplitude
$\amplitude_1/\wavevector=1$ and turning-mode amplitude
$\amplitude_2/\wavevector$ as labeled, for $\wavevector
\wormcoord\onp_1=3 \pi /2$ and $\wavevector \Delta
\wormcoord\onp=0.6$.  The turn angle $\turnanglePlanar$ and
mode-switching points $\wormcoord\onp_1$ and $\wormcoord\onp_2$
are defined in the top inset.  Bottom inset shows the dependence
of the turn angle $\turnanglePlanar$ (in degrees) on the turning-mode
amplitude $\amplitude_2 / \wavevector$. The black dots correspond
to the amplitudes of the trajectories shown.
\label{slipless turn trajectory}
}
\end{figure}
}
\section{Geometrical description of the nematode gait}
\label{crawling section}

\sloppypar \Celegans\ crawling on or burrowing through a soft gel-like
substrate (such as agar) often move with no transverse slip with
respect to the environment.  On a moist porous surface, slip may be
hindered by interfacial forces associated with water meniscus
surrounding the nematode body; in a \threed\ gel-like medium,
transverse displacements are prevented by a nonzero yield stress.

In the absence of transverse slip, all segments of the nematode body
follow a single trail, which is completely determined by the geometry
of the crawling or burrowing gait.  Thus, for a given sequence of
nematode movements, the trajectory does not depend on the substrate
mechanics.  Here we focus on this purely geometrical no-slip case.

In Sec.\ \ref{Shape and
  motion in two-dimensional space} we summarize our previously
reported PHC model
\cite{Padmanabhan-Khan-Solomon-Armstrong-Rumbaugh-Vanapalli-Blawzdziewicz:2012}
for describing body postures and trajectories of \celegans\ moving in
two dimensions.  In Sec.\ \ref{crawl 3d description} we expand our
model by adding a torsion component to the \twod\ gait representation to
achieve a \threed\ generalization of the previously established PHC
approach.

Based on our earlier
analyses of \celegans\ swimming in two dimensions
\cite{Padmanabhan-Khan-Solomon-Armstrong-Rumbaugh-Vanapalli-Blawzdziewicz:2012,%
Bilbao-Wajnryb-Vanapalli-Blawzdziewicz:2013},
we assume that PHC model and its \threed\ generalization provide a
realistic representation for the swimming gait.  In
Sec.\ \ref{maneuverability in fluids} the results derived in
Sec.\ \ref{crawling section} are used to develop a comprehensive
analysis that combines the gait geometry and motion mechanics to fully
describe \threed\ maneuvers of swimming \celegans.

\subsection{Description in \twod}
\label{Shape and motion in two-dimensional space}

The body of \celegans\ is highly elongated, and therefore its postures
can be faithfully represented by the body centerline
\cite{Padmanabhan-Khan-Solomon-Armstrong-Rumbaugh-Vanapalli-Blawzdziewicz:2012,
  Stephens-Johnson-Kerner-Bialek-Ryu:2008,
  Kim-Park-Mahadevan-Shin:2011}.
In our analysis, the shape of the centerline is described by the worm
curvature function $\wormcurvature(\curvecoord,\chronos)$, where
$\chronos$ denotes time and $\curvecoord$ is the arclength coordinate
along the body.  In what follows $\curvecoord=0$ is the tail and
$\curvecoord=\wormlength$ is the head coordinate value (where
$\wormlength$ is the nematode length).  The body configuration is
fully determined by providing the curvature function $\wormcurvature$
and the position and orientation of the nematode head (or any other
reference point along the body).

A curvilinear trail followed by the body of a nematode crawling with no
transverse slip can be described using the trail curvature
$\kappa(\wormcoord)$, where $\wormcoord$ is the arclength coordinate
along the trail.  Since at any given time $\chronos$ the configuration
of the nematode body coincides with a trail segment (see
Fig.\ \ref{curves with no torsion}), the worm and trail curvature
functions are related
\eqgone{
\begin{equation}
  \wormcurvature(\curvecoord,\chronos)=
  \curvature(\curvecoord+\wavespeed
  \chronos),
  \label{curvature equation 1}
\end{equation}
}
where $\wavespeed$ is the velocity along the trail.  Body postures and
the trail followed by a nematode are constructed from the curvature
\eqref{curvature equation 1} by solving the Serret--Frenet equations
\cite{Stoker:1969}
\eqgone{
   \begin{subequations}
      \label{2D serret-frenet equations}
      \begin{alignat}{2}
         \frac{d \utanvec}{d \wormcoord}&=\phantom{-} 
            \curvature \unorvec, \\
         \frac{d \unorvec}{d \wormcoord}&=-\curvature \utanvec,
      \end{alignat}
   \end{subequations}
}
where $\utanvec$ and $\unorvec$ are mutually perpendicular unit tangent
and normal vectors.

Observations of \celegans\ swimming in water indicate that the
propagating-curvature-wave relation \eqref{curvature equation 1}
applies not only to no-slip motion [in which case Eq.\ \eqref{curvature
    equation 1} results from geometrical constraints], but  is also
  preserved with good accuracy when significant transverse slip occurs
\cite{Padmanabhan-Khan-Solomon-Armstrong-Rumbaugh-Vanapalli-Blawzdziewicz:2012}.
This behavior is a signature of the proprioceptive coupling 
\cite{Wen-Po-Hulme-Chen-Liu-Kwok-Gershow-Leifer-Butler-Fang-Yen-Kawano-Schafer-Whitesides-Wyart-Chklovskii-Zhen-Samuel:2012} 
that allows the nematode to maintain a well organized gait in
environments with diverse mechanical properties.  Based on the
similarity between the curvature wave form for crawling and swimming
nematodes we formulate a unified gait description for different media.

\gone{
In Sec.\ \ref{Harmonic-curvature representation
  of nematode gait} we describe a simple analytical model for the
nematode curvature wave \eqref{curvature equation 1}.
}

\subsubsection{PHC representation of nematode gait}
\label{Harmonic-curvature representation of nematode gait}

\Celegans\ produces a variety of body postures, ranging from shallow
undulations to highly curved $\Omega$ shapes and loop
shapes. Recently, we have demonstrated
\cite{Padmanabhan-Khan-Solomon-Armstrong-Rumbaugh-Vanapalli-Blawzdziewicz:2012}
that all these body shapes can be accurately represented using the
harmonic curvature function
\eqgone{
\begin{equation}
  \curvature(\wormcoord)=
            {\amplitude} \cos \left(
            \wavevector \wormcoord + \phase \right),
            \label{single mode curvature}
\end{equation}
}
where $\amplitude$ is the wave amplitude, $\wavevector=2\pi/\lambda$ is
the wavevector (with $\lambda$ denoting the wavelength), and $\phase$
is the phase angle.

Combining Eqs.\ \eqref{curvature equation 1}--\eqref{single mode
  curvature} yields a family of curves representative of the most
common nematode postures.  For moderate values of the dimensionless
amplitude $\amplitude/\wavevector\approx1$, typical crawling postures (\Sshaped)
and swimming postures (\Cshaped) are obtained, as illustrated in
Fig.\ \ref{curves with no torsion}.
Higher amplitudes
correspond to $\Omega$-shapes ($A/q\approx2$) and loop shapes
($A/q\approx3$,  described in 
\cite{Padmanabhan-Khan-Solomon-Armstrong-Rumbaugh-Vanapalli-Blawzdziewicz:2012}).

According to the PHC model
\cite{Padmanabhan-Khan-Solomon-Armstrong-Rumbaugh-Vanapalli-Blawzdziewicz:2012},
trails of crawling nematodes can be accurately represented by the
harmonic function \eqref{single mode curvature} with abruptly changing
wave parameters $\amplitude$, $\wavevector$, and $\phase$ between
harmonic-curvature intervals.  Such changes generate either small
fluctuations in the direction of motion or significant body
reorientation, depending on the magnitude of the parameter variation
and the mode switching point.

\subsubsection{Planar turns}
\label{Planar turns}

To investigate the turning ability of \celegans, we consider
three-mode PHC maneuvers.  More complex trajectories can be obtained
by combining three-mode turns.  For simplicity, we focus on maneuvers
in which only the amplitude of the harmonic curvature varies,
\eqgone{
\begin{equation}
     \curvature(\wormcoord)= \left\{
     \begin{matrix}
        {\amplitude_1} \cos \left(
           \wavevector \wormcoord \right), \quad
           \wormcoord < \wormcoord\onp_1, \phantom{--.} \\
        {\amplitude_2} \cos \left(
           \wavevector \wormcoord \right), \quad
             \wormcoord\onp_1 < \wormcoord < \wormcoord\onp_2, \\
          {\amplitude_1} \cos \left(
             \wavevector \wormcoord \right), \quad
             \wormcoord\onp_2 < \wormcoord, \phantom{--.} \\
       \end{matrix}
       \right.
\label{three mode curvature}
\end{equation}
}
where $A_1$ is the base crawling amplitude (with
$\amplitude_1/\wavevector\approx1$), $\amplitude_2 > \amplitude_1$ is
the turning amplitude, and $\wormcoord\onp_1$ and $\wormcoord\onp_2$
are the mode-switching points.  Planar turns can also be obtained by
tuning the wavevector and phase angle; an analysis of such turns is
presented in the \ESM\ (\esm).

Figure \ref{slipless turn trajectory} shows an example of a crawling
nematode executing a three-mode planar turn for fixed switching points
$\wormcoord\onp_1$ and $\wormcoord\onp_2$ and several values of the
normalized turning mode amplitudes $\amplitude_2/\wavevector$.  The
dependence of the turn angle $\turnanglePlanar$ (defined in the top
inset of Fig.  \ref{slipless turn trajectory}) on the turn amplitude
$\amplitude_2$ is depicted in the bottom inset of Fig.  
\ref{slipless turn trajectory}.

\figsgone{
\clearpf

\begin{figure}
\includegraphics[width=0.34\textwidth]{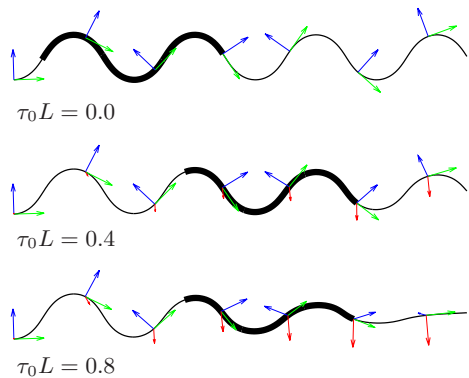}
\caption
{ ({\it Colour online}).  Curves defined by the harmonic curvature
(normalized amplitude $\amplitude / \wavevector=1$) and constant
torsion (normalized magnitude $\torsion_0 \wormlength$ as
labeled); worm contours (thick line segments) correspond to
$\wavevector \wormlength=9$.  The green, blue and red arrows are
the unit tangent, normal, and binormal vectors, respectively.
\label{curves with and no torsion}
}
\end{figure}

\begin{figure}
\includegraphics[width=\figsizeGraph]{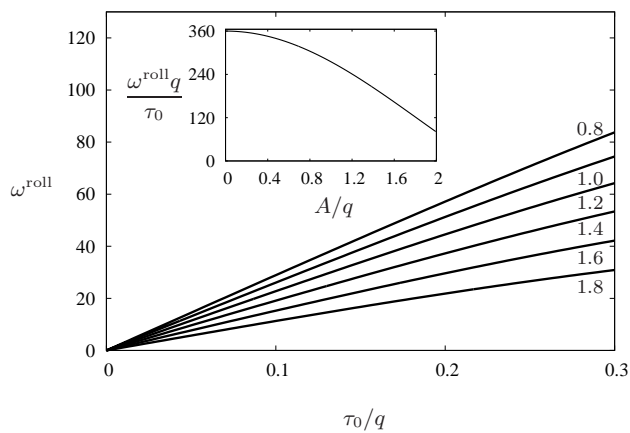}
\caption
{Roll rotation rate per undulation period $\turnvelBelly$ (in
degrees) vs normalized body torsion $\torsion_0/\wavevector$ for
one-mode torsional roll with harmonic-curvature amplitude
$\amplitude/\wavevector$ as labeled.  Inset shows the dependence
of the line slope $\turnvelBelly\wavevector/\torsion_0$ on the
amplitude $\amplitude/\wavevector$.
\label{turn rate off plane dry}
}
\end{figure}

\begin{figure}
\includegraphics[width=\figsize]{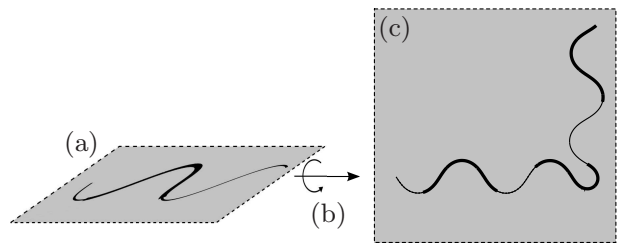}
\caption
{Nematode reorientation in \threed.  The nematode (a) undulates in
the initial plane of motion; (b) executes a torsional roll
maneuver; and (c) performs planar turn in the new plane of
motion.
\label{planes of motion exp}
   }
\end{figure}

\clearpf
}

\subsection{Description in \threed}
   \label{crawl 3d description}

\Celegans, which crawls and burrows in rotting
vegetation and swims in rainwater, performs \twod\ and
\threed\ maneuvers
\cite{Kwon-Pyo-Lee-Je:2013,%
  Voorhies-Fuchs-Thomas:2005,%
  Frezal-Felix:2015}
to find optimal environmental conditions.  Existing literature focuses
nearly exclusively on \twod\ motion, and \threed\ models of nematode
gait and locomotion mechanics are unavailable.  In this section we
expand our PHC model to incorporate \threed\ nematode postures and
trails into the gait representation.

The proposed description is motivated by recent observations of
\threed\ body postures in crawling
\cite{Deng-Xu:2014},
burrowing
\cite{Kwon-Pyo-Lee-Je:2013},
and swimming.  A crawling \celegans, for example, can produce
torsional deformation of its body, resulting in detachment of some
body sections from the underlying surface, as shown in the images of
crawling worms in Ref.\ \cite{Deng-Xu:2014}.  Swimming nematodes
undergo a similar out-of-plane body deformations, accompanied by
rapid rotation of the undulation plane (see Sec.\
\ref{maneuverability in fluids}).  Three-dimensional body actuation
seems also to occur for nictating nematodes (see supplemental movies
in
\cite{Lee-Choi-Lee-Kim-Hwang-Kim-Park-Paik-Lee:2012}).

To incorporate out-of-plane body postures into the PHC model, we use the
complete \threed\ set of Serret--Frenet equations,
\eqgone{
\begin{subequations}
      \label{full serret-frenet equations}
      \begin{alignat}{3}
         \frac{d \utanvec}{d \wormcoord}&=\phantom{-} 
            \curvature \unorvec, \\
         \frac{d \unorvec}{d \wormcoord}&=-\curvature \utanvec+
            \torsion \ubinvec, \\
         \frac{d \ubinvec}{d \wormcoord}&=-\torsion \unorvec,
      \end{alignat}
\end{subequations}
}
which includes both the curvature and torsion components.  Here
$\torsion=\torsion(\wormcoord)$ is the torsion of the curve defining
the 3D nematode trail, $\utanvec$ is the tangent unit vector,
$\unorvec$ is the normal unit vector in the local undulation plane,
and $\ubinvec=\utanvec \times \unorvec$ is the binormal unit vector
(normal to the local undulation plane).  The curve torsion $\torsion$,
which corresponds to the rotation of the Serret--Frenet basis
$(\utanvec,\unorvec,\ubinvec)$ about the tangent direction $\utanvec$,
results in out-of-plane deformations of the curve.

\subsubsection{Constant-torsion assumption}

Quantitative experimental information regarding out-of-plane body
postures of \celegans\ is very limited.  However, a qualitative
examination of nematode images  suggests that key features of
\threed\ nematode gait can be reproduced by combining the harmonic
curvature model \eqref{single mode curvature} with the
constant-torsion assumption,
\eqgone{ \begin{equation} \torsion(s)=\torsion_0.  \label{constant
torsion} \end{equation} }

The solution of the Serret--Frenet equations \eqref{full
serret-frenet equations} with the curvature and torsion given by
equations \eqref{single mode curvature} and \eqref{constant torsion}
is depicted in Fig.\ \ref{curves with and no torsion}.  The results
indicate that the general shape of the body wave is preserved in the
presence of nonzero torsion, but the undulation plane rotates with
the angular velocity proportional to the torsion amplitude
$\torsion_0$.  This behavior is in a qualitative agreement with the
crawling nematode image shown in Ref.\ \cite{Deng-Xu:2014}.  The gait
described by Eqs.\ \eqref{single mode curvature} and \eqref{constant
torsion} will be called a single-mode torsional roll.

Figure \ref{turn rate off plane dry} shows the torsional-roll rotation
rate per undulation  period,
\eqgone{
\begin{equation}
  \turnvelBelly= \cos^{-1}
  \left[
    \ubinvec(0)\bcdot\ubinvec(2 \pi / \wavevector) 
    \right].
  \label{definition of turn velocity}
\end{equation}
}
The rotation rate \eqref{definition of turn velocity} is
determined from the angle between the binormal vectors at $s=0$ and
$s=2\pi/q$; it depends linearly on the dimensionless torsion
$\torsion_0/\wavevector$.  According to the results in Fig.\ \ref{turn
  rate off plane dry}, the slope of the lines $\turnvelBelly$ vs
$\torsion_0/\wavevector$ decreases with the increasing dimensionless
curvature-wave amplitude $\amplitude/\wavevector$.  The decrease stems
from the fact that only the projection of the torsion vector
$\boldsymbol{\torsion}=\torsion\utanvec$ onto the average direction of
motion contributes to the roll.

\subsubsection{Three-mode roll maneuvers}

In the case of variable torsion, we assume that the
torsion wave $\torsion_w(\curvecoord,\chronos)$ propagates backward
along the nematode body,
\eqgone{
\begin{equation}
      \torsion_w (\curvecoord,\chronos)=
         \torsion(\curvecoord+\wavespeed \chronos),
      \label{torsion equation 1}
\end{equation}
}
similar to the curvature wave \eqref{curvature equation 1}.  For
no-slip motion (e.g., burrowing in a gel), relation \eqref{torsion
  equation 1} is required by the constraints that all segments
of the nematode body follow a single trail.  By analogy with our
\twod\ results
\cite{Padmanabhan-Khan-Solomon-Armstrong-Rumbaugh-Vanapalli-Blawzdziewicz:2012},
we assume that the form \eqref{torsion equation 1} also applies to
swimming. 

According to our qualitative observations of nematode motion in
water, \celegans\ survey their environment producing roll
reorientation maneuvers at irregular intervals.  To replicate such
motions in burrowing and swimming, we introduce a three-mode roll, where
the nematode performs one-mode harmonic curvature undulations
\eqref{curvature equation 1} superposed with piecewise constant
torsion
\eqgone{
\begin{equation}
  \torsion(\wormcoord)=
  \left\{
  \begin{matrix}
    {0} , \quad
    \wormcoord < \wormcoord\offp_1, \phantom{--.} \\
               {\torsion_0} , \quad
               \wormcoord\offp_1 < \wormcoord < \wormcoord\offp_2, \\
                              {0} , \quad
                              \wormcoord\offp_2 < \wormcoord, \phantom{--.} \\
  \end{matrix}
  \right.
  \label{three mode torsional turn}
\end{equation}
}
(where $\wormcoord\offp_1$ and $\wormcoord\offp_2$ are the mode
switching points, and
$\Delta\wormcoord\offp=\wormcoord\offp_2-\wormcoord\offp_1$ is the
roll-mode length).  By combining torsional roll maneuvers with
three-mode planar turns, a nematode is capable of exploring
\threed\ space (see Fig.\ \ref{planes of motion exp} and Video
S1 in \esm).



\begin{figure}
\includegraphics[width=\figsize]{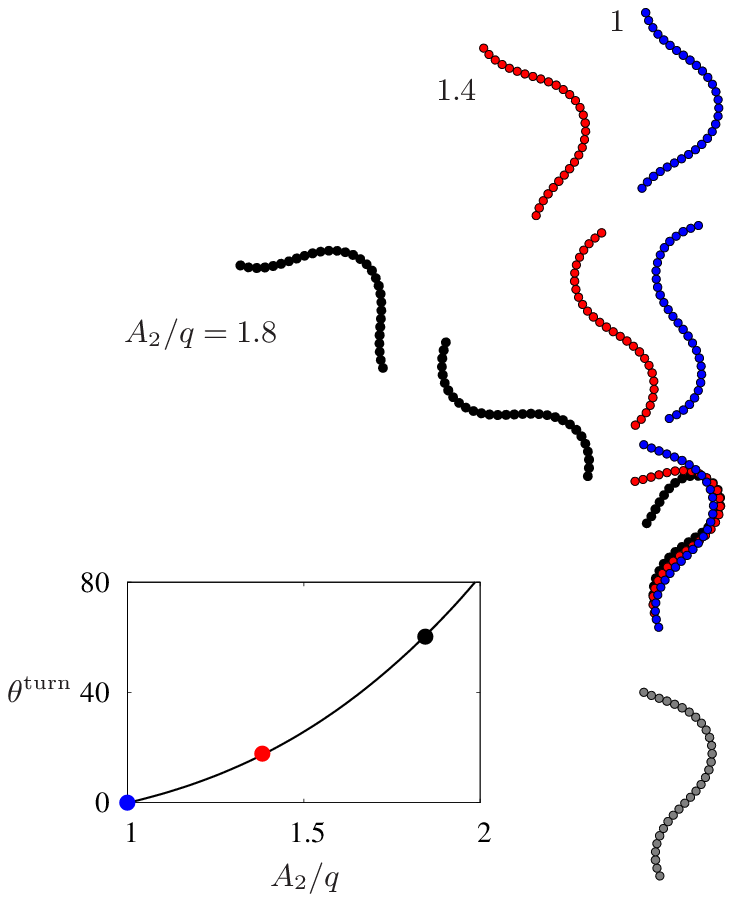}
\caption
{ ({\it Colour online}) A sequence of snapshots of a swimming
\Cshaped\ nematode ($\wavevector \wormlength=5.5$) performing a
three-mode planar turn with the default-mode amplitude
$\amplitude_1/\wavevector=1$ and turning mode amplitude as labeled,
for $\wavevector \wormcoord\onp_1=3 \pi /2$ and $\wavevector
\Delta \wormcoord\onp=0.6$.  The gray image represents the
starting position for all trajectories.  The inset shows the
dependence of the turn angle $\turnanglePlanar$ (in degrees) on the
turning-mode amplitude $\amplitude_2 / \wavevector$. The dots
correspond to the snapshots shown.
\label{on plane turn swimming}
}
\end{figure}

\begin{figure}
\includegraphics[width=\figsizeGraph]{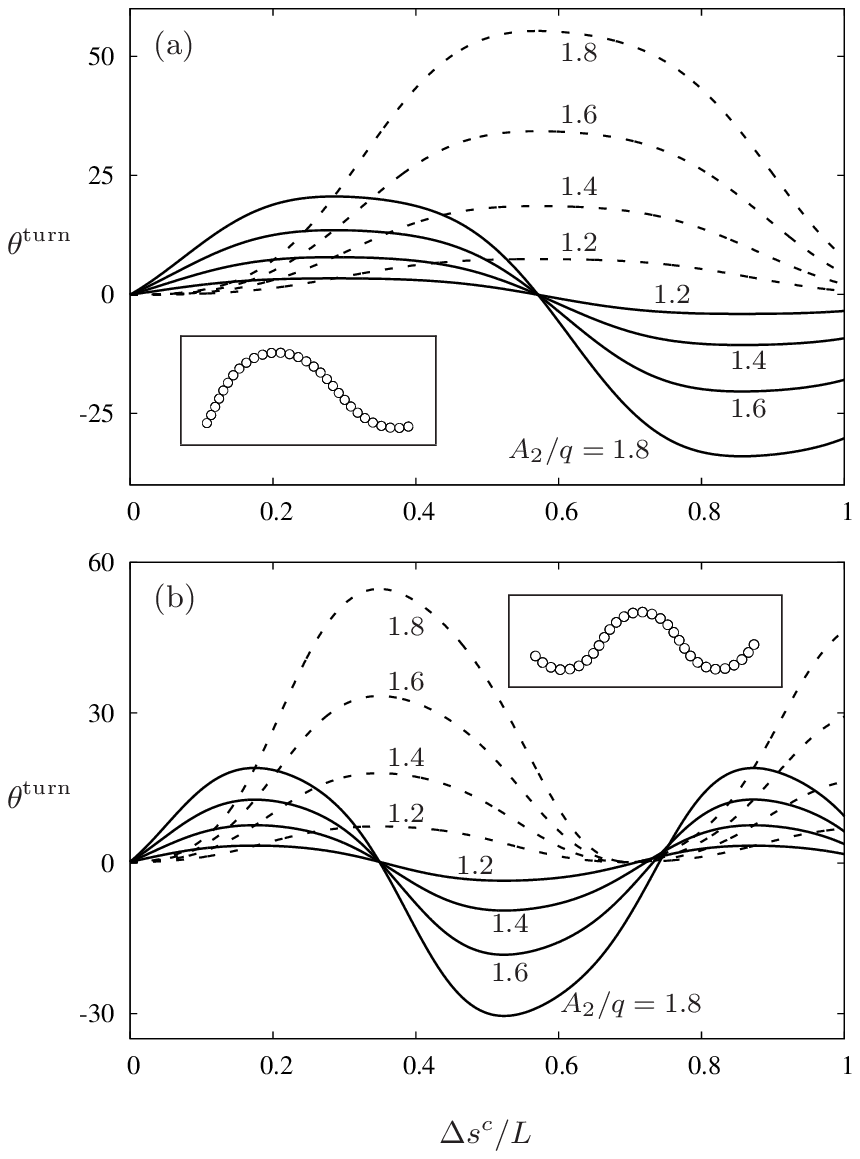}
\caption
{ Planar turn angle $\turnanglePlanar$ (in degrees) vs normalized
turning-mode length $\Delta \wormcoord\onp / \wormlength$ for (a)
\Cshaped\ ($\wavevector \wormlength=5.5$) and (b)
\Sshaped\ ($\wavevector \wormlength=9$) nematodes (see insets)
and turning-mode amplitude $\amplitude_2 / \wavevector$ as
labeled. Turns initiated at $\wavevector \wormcoord\onp_1=0$
(solid lines) and $\frac{3}{2} \pi$ (dashed). The normalized
amplitude of the default forward mode is $\amplitude_1 /
\wavevector=1$.
\label{planar turning performance of swimming worm rpy}
}
\end{figure}

\section{Nematode maneuverability in fluids}
   \label{maneuverability in fluids}

The PHC and PHC/PCT gait representations developed in
Sec.\ \ref{crawling section} will now be combined with a hydrodynamic
bead-based model to describe body reorienting maneuvers performed by
swimming \celegans.
To account for hydrodynamic forces, the nematode body is approximated
by an active chain of hydrodynamically coupled spheres arranged along
the body centerline described by Eqs.\ \eqref{three mode curvature} and
\eqref{three mode torsional turn} (see simulation images shown in
Fig.\ \ref{on plane turn swimming}).

The hydrodynamic forces and
torques acting on the chain are calculated in the Stokes-flow limit
using the generalized Rotne--Prager approximation
\cite{Wajnryb-Mizerski-Zuk-Szymczak:2013}.
This approximation is much faster (although less accurate) than the
\textsc{Hydromultipole} algorithm
\cite{Cichocki-Felderhof-Hinsen-Wajnryb-Blawzdziewicz:1994}
used in our previous study
\cite{Bilbao-Wajnryb-Vanapalli-Blawzdziewicz:2013}.  

Our methods for calculating the hydrodynamic friction and mobility
tensors are described in
Ref.\ \cite{Bilbao-Wajnryb-Vanapalli-Blawzdziewicz:2013} and in \esm.
We note that in an independent study Berman \textit{et al.}
\cite{Berman-Kenneth-Sznitman-Leshansky:2013}
also used a hydrodynamic bead-based model to investigate undulatory
swimming of a nematode.

\subsection{Swimming in \twod}
\label{Swimming in 2D}

To assess \twod\ reorientation ability of swimming \celegans, we
consider trajectories of a nematode performing three-mode PHC turns,
and evaluate the angle $\turnanglePlanar$ between the initial and
final swimming directions.  Similar to our analysis of no-slip turns
(Sec.\ \ref{Planar turns}), the nematode initially swims using the
default forward-locomotion mode with the amplitude
$\amplitude/\wavevector=1$, switches to a higher-amplitude turning
mode, and returns to the default mode.  Due to hydrodynamic effects,
the turn angle in swimming depends not only on the curvature-wave and
torsion-wave parameters, but also on the nematode length
$\wormlength$.

A comparison of the numerical simulations depicted in Fig.\ \ref{on
  plane turn swimming} with the corresponding results for a nematode
crawling without sidewise slip (Fig.\ \ref{slipless turn trajectory})
indicates that a swimming nematode produces shallower turns than its
crawling counterpart.  Moreover, due to the nonlinear dependence of
the hydrodynamic interactions on the body posture, in swimming the
relation between the turn angle $\turnanglePlanar$ and the amplitude
of the turning mode $\amplitude_2/\wavevector$ is nonlinear (in
contrast to the linear dependence for no-slip motion).

Figure \ref{planar turning performance of swimming worm rpy} provides
further details regarding the dependence of the turn angle
$\turnanglePlanar$ on the turning-mode parameters.  The results, for
different values of the turning mode amplitude, are shown vs
turning-mode length $\Delta \wormcoord\onp/\wormlength$ for two
wavevector values: $\wavevector\wormlength=5.5$ and
$\wavevector\wormlength=9$ (i.e., for \Cshaped\ and \Sshaped\ gaits).
In the scaling used in Fig.\ \ref{planar turning performance of
  swimming worm rpy}, i.e., with the mode length $\Delta
\wormcoord\onp$ normalized by the worm length $L$, the results differ
significantly between the two values of $\wavevector\wormlength$.
We note, however, that the dependence of the turn angle on the gait parameter
$\wavevector\wormlength$ is much weaker when the $\wormcoord$ coordinate
is normalized by the wavevector $\wavevector$
\cite{Bilbao-Wajnryb-Vanapalli-Blawzdziewicz:2013},
which indicates that the nematode length has only a moderate effect on
the turn dynamics.

According to the results shown in Figs.\ \ref{on plane turn swimming}
and \ref{planar turning performance of swimming worm rpy}, a swimming
nematode can control the turn angle by varying the turning-mode
amplitude $\amplitude$ as well as the turn initiation point $\wormcoord\onp_1$ and turning-mode length $\Delta\wormcoord\onp$.
Similar to the crawling case, a swimming nematode can fully explore
\twod\ space by performing a sequence of elementary turns.  For a
qualitative comparison of a \twod\ multi-mode turn of a swimming
wild-type nematode with a corresponding numerical simulation using
the PHC model see video S2 in \esm.

\figsgone{
\clearpf

\begin{figure}
\includegraphics[width=\figsize]{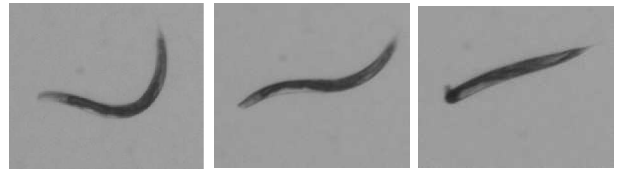}
\caption
{ Time-lapse images of a swimming \celegans\ switching the plane of
undulations.
\label{real celegans turning 3d}
}
\end{figure}

\begin{figure}
\includegraphics[width=0.33\textwidth]{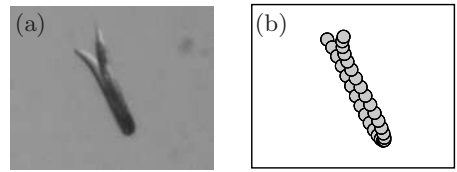}
\caption
{ (a) Non-planar body posture of a swimming \celegans\ and (b)
qualitatively matching calculated body shape with constant
torsion $\torsion_0\wormlength=0.9$ and harmonic curvature
of the normalized amplitude $\amplitude/\wavevector=1$ and
wavevector $\wavevector\wormlength=5.5$.
\label{real versus simulated worms 3d turn comparison}
} 
\end{figure}

\begin{figure}
\includegraphics[width=\figsize]{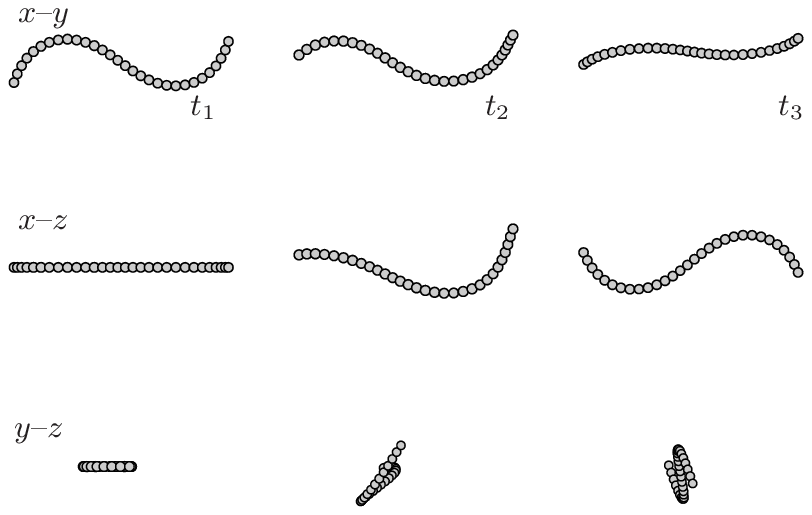}
\caption
{Stereoscopic views of a \Cshaped\ nematode
($\wavevector\wormlength=5.5$, $\amplitude/\wavevector=1$)
performing a three-mode swimming roll maneuver (projection
planes as labeled).  At times $\chronos_1$ and $\chronos_3$
the nematode performs planar undulations; at $\chronos_2$ it
performs a roll with normalized torsion
$\torsion_0\wormlength=1.5$ and length
$\wavevector\Delta\wormcoord\offp=3.24$.  The roll results in
a change of the undulation plane by more than $90^\circ$.  
\label{off plane turn stereoscopic}
}
\end{figure}

\begin{figure}
\includegraphics[width=\figsizeGraph]{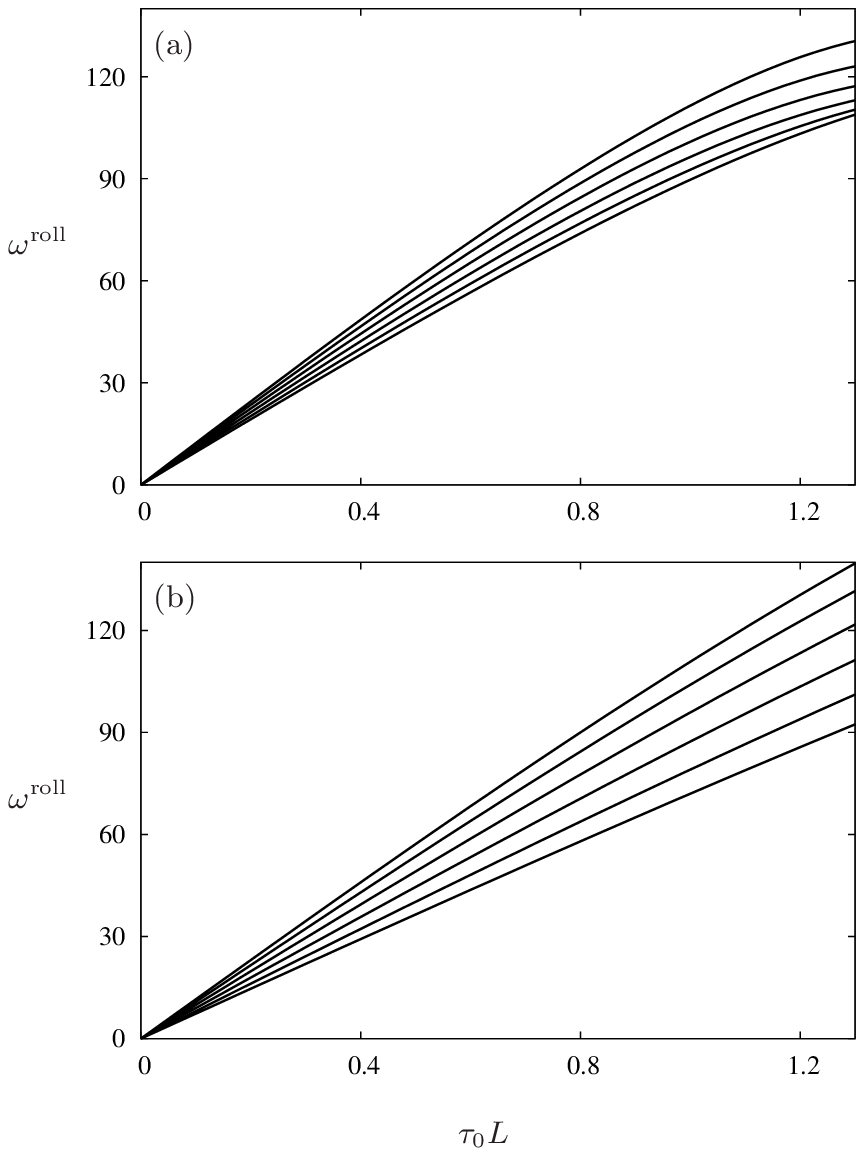}
\caption
{ Roll rotation rate per undulation period $\turnvelBelly$ (in
degrees) vs normalized body torsion $\torsion_0 \wormlength$, for
normalized curvature-wave amplitude $\amplitude / \wavevector=0.8,
1, 1.2, 1.4, 1.6, 1.8$ (from above) for (a) \Cshaped\ (\wavevector
$\wormlength=5.5$) and (b) \Sshaped\ (\wavevector $\wormlength=9$)
swimming nematodes.  \label{turning performance of swimming worm rpy}
}
\end{figure}

\begin{figure*}
\includegraphics[width=\textwidth]{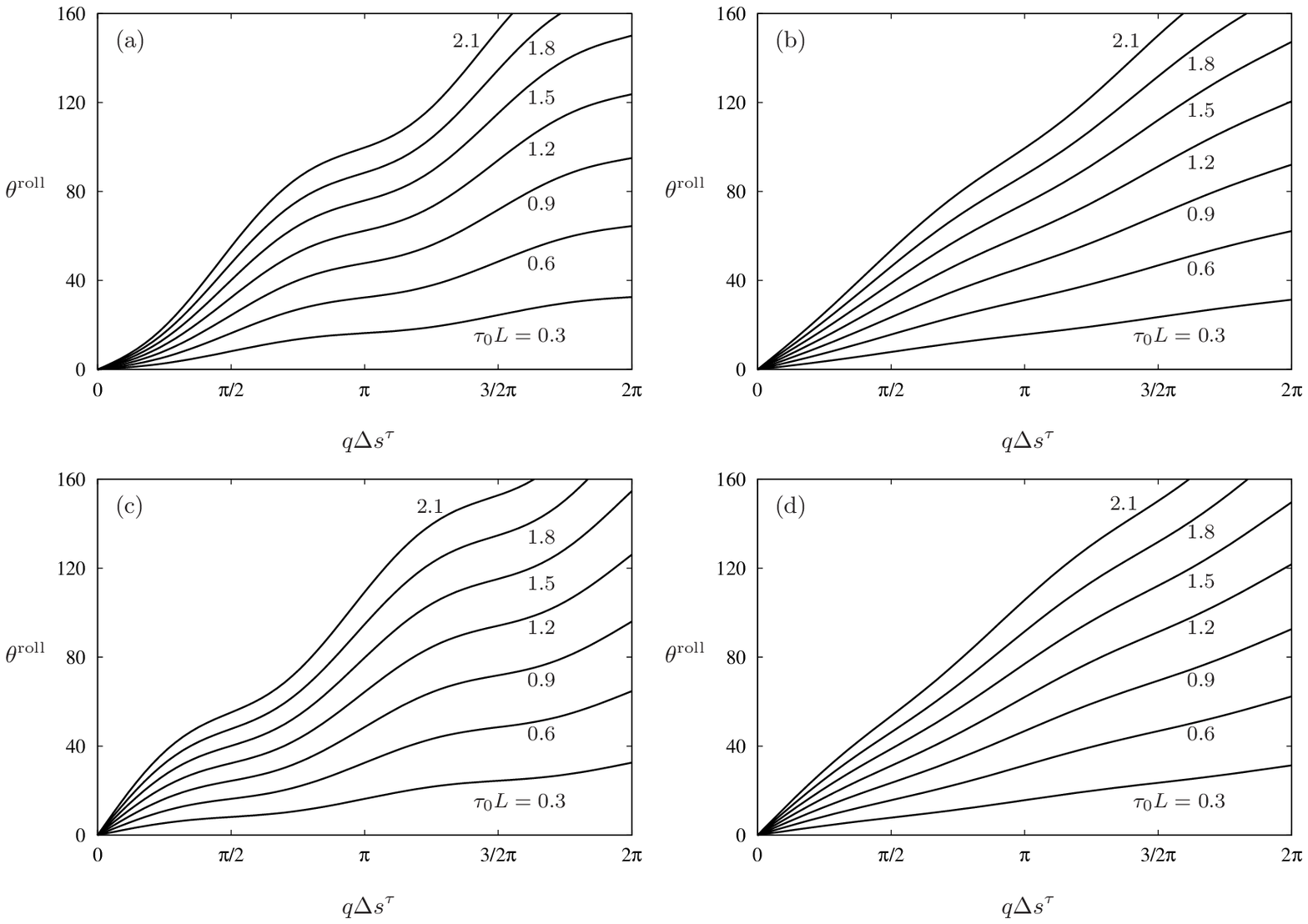}
\caption
{Roll reorientation angle $\turnangleBelly$ vs normalized
torsional mode length $\wavevector\Delta \wormcoord\offp$.
Panels (a) and (c) show results for \Cshaped\ gait
($\wavevector\wormlength=5.5$), and panels (b) and (d) show
results for \Sshaped\ ($\wavevector \wormlength=9$) gait.
Normalized magnitude of body torsion $\torsion_0 \wormlength$
as labeled.  The initial mode-switching point occurs when
the curvature at the head position is at a maximum [panels (a)
and (b)] and at a minimum [panels (c) and (d)].
\label{off plane turns of swimming worm rpy}
}
\end{figure*}

\begin{figure}
\includegraphics[width=\figsizeGraph]{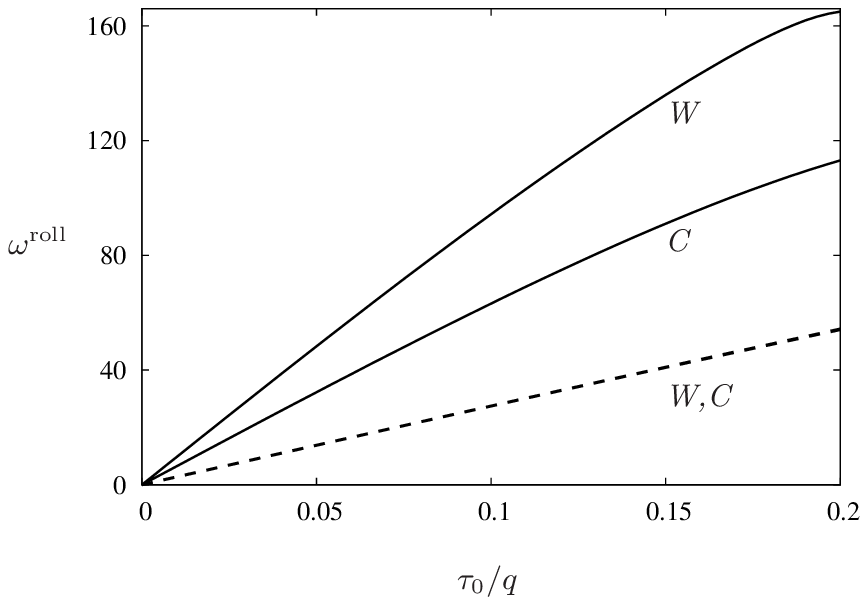}
\caption { Roll rotation rate per undulation period $\turnvelBelly$
(in degrees) vs normalized magnitude of body torsion
$\torsion_0/\wavevector$ for nematode performing one-mode rolls
for swimming (solid lines) and burrowing (dashed lines).  Results
correspond to $\amplitude/\wavevector=1$ for \Cshaped\ and
\Sshaped\ locomotive gaits, as labeled.  
\label{comparison crawling swimming turn} } 
\end{figure}

\begin{figure}[t]
\includegraphics[width=\figsize]{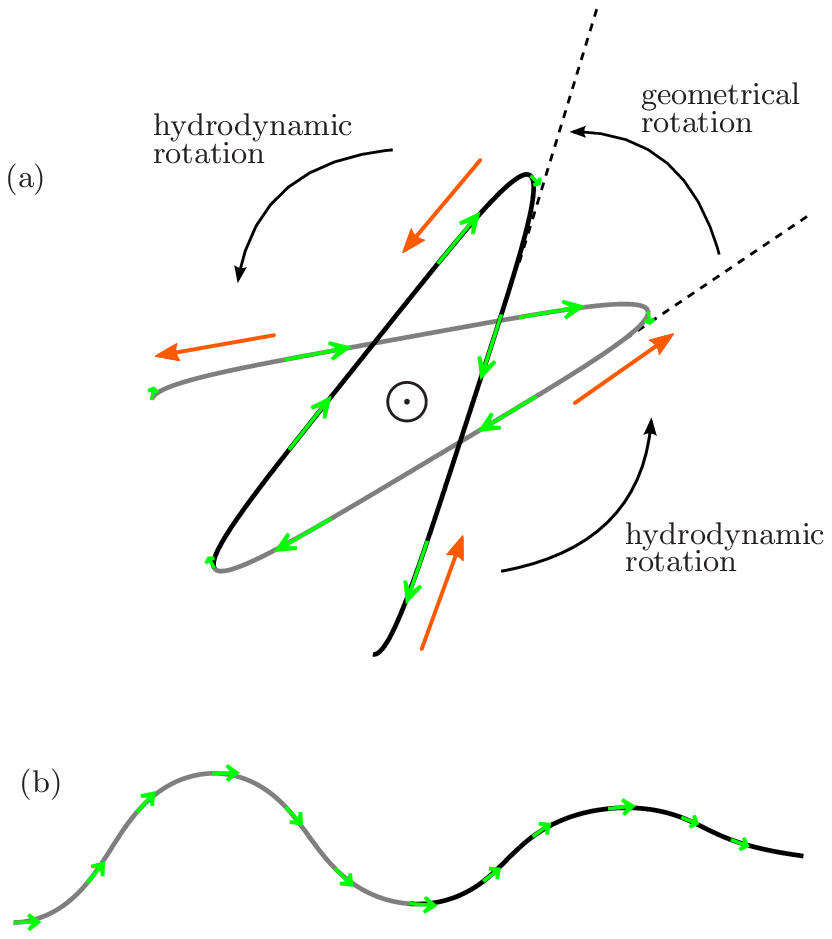}
\caption
{ (a) Front and (b) side views of a trajectory reproduced from
sinusoidal curvature with $\amplitude/\wavevector=1$ and constant
torsion $\torsion/\wavevector=0.15$.  The first period of
undulation is shown in gray and the second in black. The symbol
$\odot$ denotes the direction of propagation of the curve (out of
the page) and the green arrows represent the unit tangent vector
$\utanvec$ (in the direction of nematode motion).  In the
presence of a fluid, the nematode moving along the curve
experiences hydrodynamic resistance force (represented by orange
arrows) oriented in the same direction as the geometrical
rotation of the undulation plane.
\label{explanation for drywet comparison}
}
\end{figure}

\clearpf
}

\subsection{Swimming in \threed}
\label{offplane turns swimming}

Observations of \celegans\ swimming in bulk fluids show that the
animal is able to actively explore \threed\ environments.  A swimming
nematode can perform a series of planar turns and is also capable of
rapidly changing the plane of undulations (see Fig.\ \ref{real
  celegans turning 3d} and Video S3 in \esm).

Video images of \celegans\ undergoing undulation-plane reorientation
indicate that during such maneuvers the nematode body postures are
distorted out of the undulation plane.  We find that these nonplanar
postures resemble the torsional-roll contours defined by curvature and
torsion relations \eqref{single mode curvature} and \eqref{constant
  torsion} (see the contours depicted in Fig.\ \ref{curves with and no
  torsion}).

Figure \ref{real versus simulated worms 3d turn comparison} shows a
direct comparison between a video image of a swimming nematode and a
calculated torsional-roll posture with a curvature wavevector
$\wavevector\wormlength=5.5$ (\Cshaped\ gait) and constant torsion
$\torsion_0\wormlength=0.9$.  We note that the out-of-plane
deformation of the nematode body is quite small for this torsion
value; our hydrodynamic analysis shows, however, that even a moderate
torsional deformation may have a significant dynamic effect, producing
rolls that are significantly stronger than the corresponding rolls in
no-slip burrowing.

To elucidate the effect of the torsional roll in swimming, we present
results of our hydrodynamic simulations for single-mode and three-mode
roll maneuvers, described by Eq.\ \eqref{constant torsion} and
\eqref{three mode torsional turn}, respectively.  The overall
undulation-plane rotation rate for a swimming nematode includes
geometrical and hydrodynamic contributions.  The geometrical component
results from the rotation of the Serret--Frenet frame
$(\utanvec,\unorvec,\ubinvec)$ according to equations \eqref{full
  serret-frenet equations}; the hydrodynamic component is associated
with the rigid-body rotation of the whole nematode body, as required
by the zero net hydrodynamic force and torque conditions.

Figure \ref{off plane turn stereoscopic} shows stereoscopic snapshots
of the nematode configuration at the beginning, during, and at the end
of a simulated three-mode roll. Another example of a three-mode roll
maneuver is presented in Video S4 in \esm.  In both cases the nematode
undergoes a roll rotation of approximately $90^{\circ}$.  
Simulated nematode motion in the Video S4 closely resembles the
behavior of a wild-type nematode seen in Video S3.

Quantitative results regarding the effectiveness of roll reorientation
maneuvers in fluids are presented in Figs.\ \ref{turning performance
  of swimming worm rpy}--\ref{comparison crawling swimming turn}.  The
roll rotation rate per undulation period \eqref{definition of turn
  velocity} during a single-mode roll is plotted in Fig.\ \ref{turning
  performance of swimming worm rpy}, and the total roll reorientation
angle
\eqgone{
\begin{equation}
  \turnangleBelly=\cos^{-1}
  \left[
    \ubinvec(\wormcoord\offp_1) \bcdot
    \ubinvec(\wormcoord\offp_2)
    \right]
  \label{definition of offplane turn angle}
\end{equation}
}
for a nematode performing a three-mode roll is plotted in
Fig.\ \ref{off plane turns of swimming worm rpy}. 

The roll rotation rate $\turnvelBelly$ depicted in Fig.\ \ref{turning
  performance of swimming worm rpy} is shown vs the dimensionless
torsion amplitude $\torsion_0\wormlength$, for several values of the
normalized curvature amplitude $\amplitude / \wavevector$, for
\Cshaped\ and \Sshaped\ gaits.  Note that in swimming the roll
rotation rate depends on the nematode length whereas in no-slip
burrowing $\turnvelBelly$ depends only on the torsion and curvature-wave
parameters.  Hence a different normalization is used in
Figs.\ \ref{turning performance of swimming worm rpy} (swimming) and
\ref{turn rate off plane dry} (burrowing).

In swimming, similar to the results for the no-slip roll
(Fig.\ \ref{turn rate off plane dry}), the roll rotation rate is an
increasing function of the normalized curvature-wave amplitude
$\amplitude/\wavevector$ and body torsion $\torsion_0\wormlength$, but
the dependence on the torsion is nonliner due to hydrodynamic effects.
The results for \Cshaped\ and \Sshaped\ gaits have a similar magnitude
for given values of $\amplitude/\wavevector$ and $\torsion_0
\wormlength$.

The total reorientation angle $\turnangleBelly$ during a three-mode
roll maneuver is plotted in Fig.\ \ref{off plane turns of swimming
  worm rpy} vs the dimensionless torsional mode length
$\wavevector\Delta\wormcoord\offp$, normalized by the wavevector of
the curvature undulations.  The results are similar for the
\Cshaped\ and \Sshaped\ gaits and are only moderately affected by the
initial mode switching point.  The magnitude of the reorientation
angle $\turnangleBelly$ is consistent with the single-mode roll
rotation rate presented in Fig.\ \ref{turning performance of swimming
  worm rpy}.

Our experimental observations of \celegans\ swimming maneuvers
indicate that the nematode typically needs one to two periods of
motion to roll by $90^\circ$, and rolls by $90^\circ$ in less than one
period are not unusual.  According to the results shown in
Figs.\ \ref{turning performance of swimming worm rpy} and \ref{off
  plane turns of swimming worm rpy}, a $90^\circ$ roll corresponds to
the body torsion of $\torsion_0\wormlength\approx1$ for the roll mode
length ranging from half to one undulation period.  Shorter modes
require larger torsion (see, e.g.\ Video S4 in \esm).

\subsubsection{Hydrodynamic enhancement in swimming rolls} 

To compare the swimming roll rotation rate $\turnvelBelly$ shown in
Fig.\ \ref{turning performance of swimming worm rpy} with the
corresponding purely geometrical no-slip result 
(Fig.\ \ref{turn rate off plane dry}), the swimming and no-slip
simulation data are replotted in Fig.\ \ref{comparison crawling
  swimming turn} using the same scaling of the torsion
$\torsion_0/\wavevector$ for both data sets.  The results indicate
that hydrodynamic forces in swimming can enhance the rotation rate by
a factor ranging from approximately two (\Cshaped\ gait) to even
three (\Sshaped\ gait).  While this hydrodynamic enhancement may
seem surprising, its origin can be explained by a careful examination
of hydrodynamic forces acting on a nematode performing a torsional
roll in the absence of sidewise slip.

The projection of a torsional no-slip trail onto the plane normal to
the overall direction of motion is depicted in Fig.\ \ref{explanation
  for drywet comparison}.  The graph  shows that the plane of
undulations gradually rotates anticlockwise, whereas the nematode
moves clockwise around the axis of the trajectory.  In fluids, this
clockwise motion produces the oppositely oriented hydrodynamic
resistance, which causes hydrodynamically induced nematode rotation in
the anticlockwise direction.  Since the geometrical and hydrodynamic
contributions to $\turnvelBelly$ have the same sign, they add up,
resulting in the enhanced reorientation.

We believe that swimming \celegans\ use the above hydrodynamic
enhancement mechanism in the observed rapid roll maneuvers (see Video
S3 in \esm).  Our numerical simulations presented in Video S4 in \esm\ 
reproduce this behavior.


\figsgone{
  \clearpf

\begin{figure}
\includegraphics[width=\figsize]{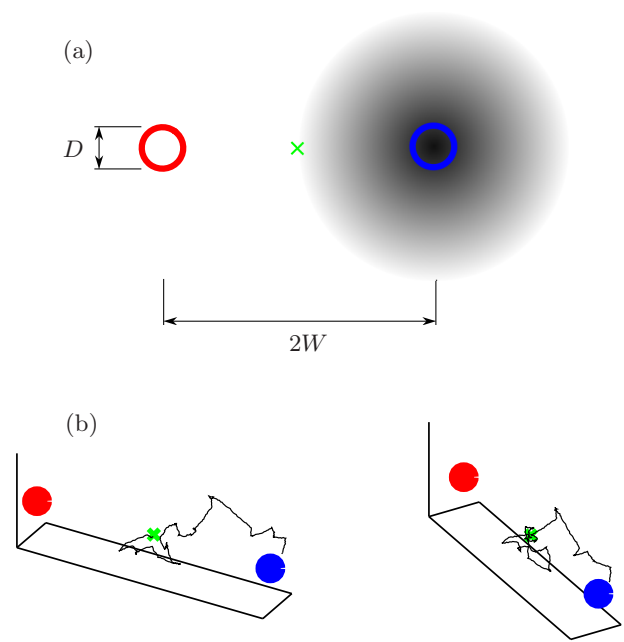}
\caption
{ (\textit{Colour online}) (a) Top view of the simulated chemotaxis
assay in \threed.  Nematodes are initialized with a random
orientation at the green symbol $\times$ located midway between
the position of a Gaussian chemoattractant concentration peak
(right) and a spherical control region represented by the left
red circle. The right blue circle represents a spherical test
region.  (b) Two views of a simulated path followed
by a chemotaxing \celegans\ in a \threed\ fluid.
\label{chemotaxis apparatus}
}
\end{figure}

\begin{figure}
\includegraphics[width=\figsizeGraph]{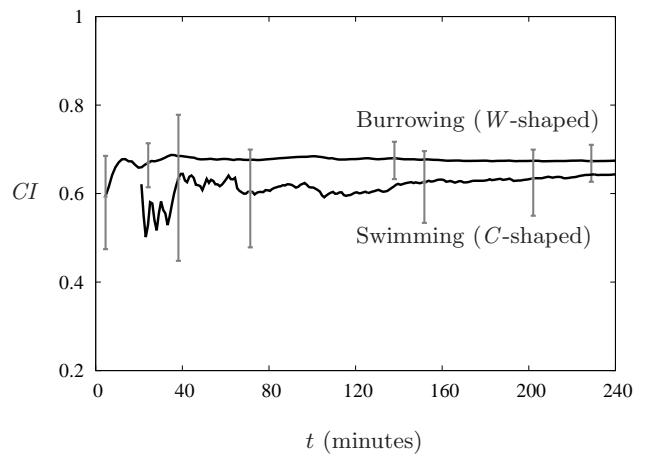}
\caption
{ Chemotaxis index $\Ci$ vs time for burrowing and swimming
nematodes in \threed.  The sensing parameters for the
burrowing and swimming cases are the same, but the undulation
frequency of the swimmer is four times larger than the
undulation frequency of the burrower.
\label{chemotaxis saturation}
}
\end{figure}

\begingroup
  \begin{table*}
  \begin{ruledtabular}
   \begin{tabular}{ l l l }
   Parameter & Name & Value \\
   \hline
   $\turnsensitivity^\prime$ & Chemoattractant sensitivity parameter& $5\,\textrm{s}/\textrm{mm}$ \\
   $\turnrateconst$ & Turn initiation rate&$10\textrm{ turns}/\textrm{min}$ \\
   $\turnratehigh$ &Upper turn-rate saturation limit& $15\textrm{ turns}/\textrm{min}$ \\
   $\turnratelow$ &Lower turn-rate saturation limit& $5\textrm{ turns}/\textrm{min}$ \\
   $\corkrateconst$ & Roll initiation rate  & $37\textrm{ rolls}/\textrm{min}$ \\
   $\decay$ & Memory time constant & $1.6\,\textrm{s}^{-1}$ \\
   $\amplitude_1/\wavevector$ & Forward mode normalized amplitude & 1 \\
   $\frequency_\burrowing$ & Burrowing gait frequency & $0.44\,\textrm{s}^{-1}$ \\
   $\frequency_\swimming$ & Swimming gait frequency & $1.75\,\textrm{s}^{-1}$ \\
   $\wavevector_\burrowing\wormlength$ & Burrowing gait normalized wavevector & $9$ \\
   $\wavevector_\swimming\wormlength$ & Swimming gait normalized wavevector & $5.5$ \\
   $\wormlength$ & Worm length & $1\,\textrm{mm}$\\
   \end{tabular}
   \end{ruledtabular}
   \caption
   {
      \label{chemosensing parameters}
      System parameters for \threed\ simulations of chemotaxis of
      burrowing and swimming \celegans.  The chemoattractant
      sensitivity parameter $\turnsensitivity^\prime=\turnsensitivity/
      \nabla \concentration|_{r=\gaussWidth}$ is normalized by the
      maximum chemical concentration gradient located at
      $r=\gaussWidth$.  The memory time constant $\decay$ corresponds
      to an approximately $5\,\textrm{s}$ half-decay response time of
      sensory neurons to the step increase of chemoattractant
      concentration
      \cite{Suzuki-Thiele-Faumont-Ezcurra-Lockery-Schafer:2008}.}
  \end{table*}
  \endgroup

     \clearpf
}
     
\section{Application to Chemotaxis}
   \label{chemotaxis in 3d}

According to the locomotion model developed in Secs.\ \ref{crawling
  section} and \ref{maneuverability in fluids}, \celegans\ navigates
\threed\ environments by combining forward motion, planar turns, and
torsional rolls (see Videos S1 and S5 in \esm\ for simulations of
burrowing and swimming behavior).  Here we
show that by modulating the rate of the above elementary
maneuvers in response to a chemoattractant concentration variation,
\celegans\ can efficiently chemotax in 3D.

It has been demonstrated that in \twod\ environments \celegans\ employs
two chemotaxis mechanisms: klinotaxis (gradual turn) where the
nematode actively reorients itself towards the concentration gradient
direction \cite{Iino-Yoshida:2009}, and klinokinesis (biased random
walk), where it adjusts time intervals of motion in favorable and
unfavorable directions to achieve a net displacement towards the
higher concentration \cite{Pierce-Shimomura-Morse-Lockery:1999}.

While we expect both strategies to be effective in \threed, in the
present paper we focus on the biased random walk (BRW) mechanism.  A
more comprehensive study of nematode chemotaxis in complex environment
will be described elsewhere.

\subsection{Chemical sensing model} 

\label{chemical sensing}

A moving \celegans\ assesses the spatial gradients of the
chemoattractant concentration by monitoring its time variation
$\concentration(\chronos)$ at the current position of chemical
receptors (located in the nematode head segment) \cite{Lockery:2011}.  To
model this process we use an approach that has been successfully
applied to describe bacterial chemotaxis
\cite{Celani-Vergassola:2010}.

Accordingly, we assume that the rate of turns is modulated by the
chemical signal
\eqgone{
\begin{equation}
  \csignal(\chronos)=\int_0^\chronos
  \concentration(\chronos-\dtime)
  \response(\dtime) d \dtime,
  \label{chemo signal}
\end{equation}
}
where $\response(\chronos)$ is the memory function that describes the
response of sensory neurons to chemoattractant.  Assuming that the
nematode responds to changes of the concentration of chemoattractant
but not to the absolute concentration value \cite{Lockery:2011}, we
impose the ideal adaptation condition \cite{Celani-Vergassola:2010}
\eqgone{
\begin{equation}
  \int_0^\infty \response(\chronos) d \chronos=0
  \label{chemo norm 1}.
\end{equation}
}
We also assume the normalization 
\eqgone{
\begin{equation}
  \label{normalization of M}
\int_0^\infty \chronos\response(\chronos) d \chronos=1.
\end{equation}
}

In our numerical calculation we use the memory function defined by the
expression 
\eqgone{
\begin{equation}
  \response(\chronos)=\decay^2 \mathrm{e}^{\decay \chronos}
  \left[
    (\decay \chronos)-\frac{1}{2}(\decay \chronos)^2
    \right],
  \label{response function}
\end{equation}
}
where $\decay^{-1}$ is the decay time \cite{Celani-Vergassola:2010}.
The memory function \eqref{response function} is positive for
$0<\chronos<\decay^{-1}$ and negative for $\chronos>\decay^{-1}$;
the positive and negative regions have equal integrals.
The convolution integral \eqref{chemo signal} filters out short-time
fluctuations of the chemoattractant concentration and provides
information whether the concentration is increasing ($\csignal>0$) or
decreasing ($\csignal<0$) on the timescale $\decay^{-1}$.

\subsection{Biased random walk}

\label{biased random motion}

A nematode performing BRW generates a trajectory drift towards a peak
of chemoattractant concentration by controlling the frequency of
random turns in response to sensory signals.  Namely, to achieve
longer intervals of motion towards the peak, the turn rate is reduced
(increased) when the animal moves towards (away from) the peak.  In
our chemotaxis model, the chemical signal $\csignal$ controls the
frequency of elementary three-mode turns.

The nematode burrows or swims using the default 
harmonic-curvature locomotion mode with
$\amplitude/\wavevector\approx1$, and switches randomly to a turning
mode at a turn initiation rate $\turnrate(\chronos)$, which depends on
the chemotaxis signal \eqref{chemo signal}.  The nematode also
performs roll maneuvers initiated at a rate $\corkrate(\chronos)$.

To mimic the experimentally observed sigmoidal character of the turn
response to the chemoattractant concentration variation
\cite{Pierce-Shimomura-Morse-Lockery:1999}, we assume that the turn
initiation rate $\turnrate$ has a linear-response region and a saturation
cutoff,
\eqgone{
\begin{equation}
  \turnrate(\chronos)=
  \left\{
  \begin{matrix}
    \hspace{-12mm} \turnratelow; \hspace{4mm} f(\csignal) \le 
    \turnratelow \\
    f(\csignal); \hspace{4mm} \turnratelow <
    f(\csignal) < \turnratehigh, \\
    \hspace{-11mm} \turnratehigh; 
    \hspace{4mm} \turnratehigh \le f(\csignal) \\
  \end{matrix}
  \right.
  \label{rate of turn}
\end{equation}
}
where
\eqgone{
\begin{equation}
  f(\csignal)=
  \turnrateconst \left[
    1 - \turnsensitivity \csignal(\chronos)
    \right].
  \label{rate of turn mid}
\end{equation}
}
Here $\turnrateconst$ is the turn rate in the absence of
chemoattractant, $\turnsensitivity$ is a chemoattractant sensitivity
parameter, and $\turnratehigh$ and $\turnratelow$ are the saturation
limits.
Since there are no experimental data regarding the effect of
chemoattractant on roll maneuvers, we assume that the roll initiation
frequency is constant and independent of chemical sensing,
\eqgone{
\begin{equation}
  \corkrate(\chronos)=\corkrateconst.
  \label{rate of cork}
\end{equation}
}

In Sec.\ \ref{Chemotaxis in 3D} we show that assumptions
\eqref{rate of turn} and \eqref{rate of cork} are sufficient to
produce efficient \threed\ chemotaxis.

\subsection{The geometry of simulated chemotaxis assay}

In our \threed\ numerical simulations of chemotaxis of burrowing and
swimming nematodes [see Fig.\ \ref{chemotaxis apparatus}(a)], the
simulated worms are placed at a designated starting point at the edge of a
Gaussian chemoattractant concentration distribution.  The chemotaxis
efficiency is determined by comparing the number of nematodes that
reach a spherical test region at the chemoattractant concentration
peak with the number of those that reach a symmetrically placed control
region in the low-concentration domain.  Our \threed\ arrangement mimics
a typical experimental setup for investigating \twod\ chemotaxis on
agar surface
\cite{%
  Bargmann-Horvitz:1991,%
  Nishino-Sato-Ito-Matsuura:2015}.

The chemotaxis efficiency is quantified using the chemotaxis index
\eqgone{
\begin{equation}
  \Ci(\chronos)=\frac{\nowormsA(\chronos)-\nowormsC(\chronos)}
           {\nowormsA(\chronos)+\nowormsC(\chronos)},
           \label{chemotaxis index cumulative}
\end{equation}
}
where $\nowormsA(\chronos)$ and $\nowormsC(\chronos)$ denote the
number of nematodes that reached the test and control regions,
respectively, during the simulation interval $[0,\chronos]$.  The
simulation of a nematode trajectory is terminated when the nematode
enters either the test or control region (in \twod\ laboratory
experiments these regions usually contain anesthetic to immobilize the
nematodes \cite{Nishino-Sato-Ito-Matsuura:2015}).  A sample trajectory
of a swimming \celegans\ reaching the test region is shown in Fig.
\ref{chemotaxis apparatus}(b).

In our simulations we consider a system with the following parameters.
The spatial concentration of chemoattractant  is given by a \threed\
Gaussian distribution
\eqgone{
\begin{equation}
  \label{Gaussian}
  \concentration\sim\exp\left(-r^2/2\variance^2\right)
\end{equation}
}
with the peak position at a distance $\dishWidth=36\wormlength$ from
the starting point and the width $\gaussWidth=0.4\dishWidth$.  The
diameter of the test and control regions is $D=10\wormlength$. For the
worm length $\wormlength=1\,\textrm{mm}$, the starting point is thus
at a distance $3.6\,\textrm{cm}$ from the chemoattractant
concentration maximum, and the test and control regions have a
$1\,\textrm{cm}$ diameter (similar to the geometry of
\twod\ experiments in a Petri dish).

\subsection{Simulation results}
\label{Chemotaxis in 3D}

We have performed simulations for two cases: (a) burrowing
\Sshaped\ nematodes (no transverse slip) and (b) swimming
\Cshaped\ nematodes.  The nematode gait parameters and parameters of
the chemotaxis model are listed in Table \ref{chemosensing parameters}
(see Sec.\ \ref{Chemotaxis simulation details} for more details).
The gait parameter values are based on observations of crawling and
swimming \celegans.  We use the same chemosensation and turn-rate
parameters in the burrowing and swimming simulations to isolate the
effect of the locomotion mechanics on the chemotaxis efficiency.

Chemotaxis index \eqref{chemotaxis index cumulative}, shown in
Fig.\ \ref{chemotaxis saturation}, is significantly greater than zero,
indicating a clear trajectory bias towards the chemoattractant
concentration peak.  At long times the value of $\Ci$ for both
burrowing and swimming is approximately the same.  However, the
burrowing nematodes are able to reach the chemoattractant peak more
quickly, due to the higher burrowing velocity (twice the swimming
velocity for the gait parameters considered).

The chemotaxis index for burrowing and swimming nematodes is similar
due to two compensating effects: (a) for swimming nematodes the
chemical signal \eqref{chemo signal} is weaker because, as a result of
their lower velocity, swimming worms detect a smaller time variation
of chemoattractant concentration; while (b) the number of turns per unit
trajectory length is larger for swimming nematodes because of their lower
velocity.  More turns implies a stronger bias towards chemoattractant,
which compensates for a weaker chemical signal.

We expect that similar compensating factors occur in other
environments as well.  Thus, the nematode may achieve efficient
chemotaxis without adjusting turn-rate control parameters in response
to a variation in mechanical properties of the medium in which it
moves; a more complex control system that relies on mechanical
feedback is unnecessary.

The values of the \twod\ and \threed\ chemotaxis index do not differ
either, but in \threed\ many more nematodes escape without reaching
either the test or control region; this effect should be taken into
account in design of a testing platform for future
\threed\ chemotaxis experiments.

\clearpf

\section{Summary and discussion}
\label{Summary and discussion}

To provide mathematical tools for analysis of \threed\ nematode
maneuvers, we have developed a gait model that
combines piecewise-harmonic-curvature (PHC) and
piecewise-constant-torsion (PCT) body-shape assumptions.  We have
demonstrated that nematode reorientation in an arbitrary direction can
be achieved by a mix of \twod\ turns (driven by changes of
curvature-wave parameters) and \threed\ roll maneuvers (associated with nonzero
body torsion).

We have applied our model to investigate \threed\ motion of nematodes
burrowing in gels and swimming in Newtonian fluids.  We examined the
effectiveness of the turns and rolls; our modeling of torsional roll
maneuvers reveals a strong enhancement of the roll-rotation rate due
to hydrodynamic forces.  This hydrodynamic effect may explain the
observed rapid reorientation of the undulation plane of swimming
\celegans\ (see video S3 in \esm).

We have used our gait representation to study nematode chemotaxis in
gels and fluids.  We have shown that the nematode does not need to
adjust its sensory-motor apparatus to effectively respond to chemical
stimuli in environments with different mechanical properties.
Chemotaxis of \celegans\ in \threed\ environments was experimentally
demonstrated
\cite{Pierce-Shimomura-Chen-Mun-Ho-Sarkis-McIntire:2008,%
  Beron-Vidal-Gadea-Cohn-Parikh-Hwang-Pierce-Shimomura:2015},
but, to our knowledge, the formulation presented here is the first
quantitative analysis that takes into account the gait geometry and
swimming mechanics.

\gone{Our numerical simulations of nematode trajectories in which the turn
parameters $\wormcoord\onp_1$ and $\Delta\wormcoord\onp$ are randomly
chosen yield many turns that produce only small directional changes.
In contrast, experiments show that most $\Omega$ postures adopted by
\celegans\ result in relatively sharp turns.  Our calculations thus
suggest that the nematode is able to choose optimal turn initiation
and termination points.}

By supplying an analytical description of elementary body movements,
our PHC/PCT gait model is applicable to analyze nematode behavior in
\threed\ complex media, including inhomogeneous materials and
non-Newtonian fluids.  We expect that our results may also shed light
on other subtle aspects of neuromuscular control of
\threed\ locomotion of \celegans.

Body movements of \celegans\ are actuated by coordinated action of
four quadrants of body muscles, which are arranged into double rows
that span the entire body length \cite{Altun-Hall:2009}.  While this
muscle anatomy is compatible with \threed\ body actuation, it is less
clear how the neural system can activate differential muscle
contractions needed to generate \threed\ body deformations out of the
dorsoventral undulation plane.

The head and neck body muscles are innervated by the nerve ring motor
neurons, which synapse onto cells in individual quadrants (and thus
are capable to actuate each quadrant individually).  In contrast, the
remaining body muscles are innervated only by the ventral-cord motor
neurons, which synapse onto muscles of either two ventral or two
dorsal quadrants, consistent with \twod\ actuation
\cite{White-Southgate-Thomson-Brenner:1986,%
Altun-Hall:2009,%
Riddle-Blumenthal-Meyer-Priess:1997}.
Based on this topology of neural connections, it is generally assumed
that only the head and neck segments can perform 3D motion.

However, the stereoscopic experiments by Kwon \textit{et al.\/}
\cite{Kwon-Pyo-Lee-Je:2013} clearly show that burrowing nematodes can
adopt a variety of \threed\ postures, in which the entire body
undergoes out-of-plane deformations.  In the swimming case the
evidence is less conclusive because stereoscopic images are not
available.  Nonetheless, two-dimensional images, such as the one shown
in Fig.\ \ref{real versus simulated worms 3d turn comparison}, agree
qualitatively with our \threed\ body posture model.

Our hydrodynamic calculations of the roll dynamics further support a
conclusion that the entire nematode body participates in generating
the roll motion.  In particular, we find that torsional wave
propagating throughout the entire body length can produce a
90${}^\circ$ roll in less than one undulation period.  In contrast, a
stroke in which only the head and neck segments are displaced out of
the undulation plane produces weaker rolls than the ones observed in
our experiments (see the results of our additional simulations in
Fig.\ S2 in \esm).

We hypothesize that the generation of a nonzero torsion in the
posterior part of the nematode body involves a \threed\ proprioceptive
feedback, which enables propagation of the torsional wave to the body
region innervated only by the ventral-cord motor neurons.  The
existence of \twod\ proprioceptive feedback in \celegans\ and its
important role in generation of a planar undulation wave have been
demonstrated
\cite{Wen-Po-Hulme-Chen-Liu-Kwok-Gershow-Leifer-Butler-Fang-Yen-Kawano-Schafer-Whitesides-Wyart-Chklovskii-Zhen-Samuel:2012}. 
We thus conjecture that \celegans\ may employ \threed\ proprioception
to generate well controlled out-of-plane body movements.  (An
alternative hypothesis would be that the worm may control its
\threed\ body movements using direct communication between muscle cells
via gap junctions.)

Current evidence is not sufficient to rule out a passive, purely
mechanical origin of propagation of \threed\ deformations along a
portion of the nematode body outside the head and neck regions.  In
particular, the body shapes observed during burrowing
\cite{Kwon-Pyo-Lee-Je:2013} can, in principle, be generated by
\threed\ actuation of the head and neck muscles, with the rest of the
body passively sliding along the tunnel carved by the nematode head.  Such
motion, however, would be inefficient.  

According to Wen \textit{et al.} 
\cite{Wen-Po-Hulme-Chen-Liu-Kwok-Gershow-Leifer-Butler-Fang-Yen-Kawano-Schafer-Whitesides-Wyart-Chklovskii-Zhen-Samuel:2012},
``[t]he cellular economy of the \celegans\ wiring diagram
implies that individual neurons may have high levels of complexity''
(for example, B-type motor neurons are able to transduce
proprioception).  We conclude that it is thus conceivable that
functional complexity on a single-neuron level allows symmetrically
connected neurons to produce differential muscle excitation.

Nature created a four-quadrant muscle arrangement that is anatomically
capable of generating \threed\ body deformations.  Therefore, our
question is: why should there not be a compatible
neuromuscular-control system that would allow the nematode to take
advantage of this (existing) muscle structure?  Further analyses of
\threed\ locomotion of \celegans\ will help determine if such a
control system exists.


\clearpf
\section{Methods}
\label{experimental}

\subsection{Imaging}

Wild type \celegans\ was used to record the 3D swimming in our
experiments.  Worms were cultured in NGM plate (Nematode Growth
Medium) seeded with bacteria \textit{E.\ coli} OP50 at
20${}^\circ$C. A liquid pool was created for swimming assay by adding
5--7\,mL of M9 buffer into a 60\,mm Petri dish. The pool height was
approximately 2\,mm. The surface of the Petri dish was treated with
5\,\% Pluoronic F127 for 5\,min before adding M9 buffer.  Individual
young adults were manually transferred into a liquid pool of M9 buffer
using a worm pick.  Swimming episodes were recorded at 5 frames per
second using an SVSI camera and Zeiss stemi 2000-C stereo microscope
imaging system at 1.6X magnification with a pixel resolution of 0.21
pixels/$\mu$m. All imaging was carried out in a food-free environment
at 20${}^\circ$C.

\subsection{Chemotaxis simulation details}
\label{Chemotaxis simulation details}

In our simulations, the planar turns and rolls are initiated randomly
at the rates $\turnrate$ and $\corkrate$, provided that a
reorientation maneuver of the same kind is not already taking place.
The turns and rolls are performed
independently.

The default gait parameters (used between reorientation maneuvers)
have fixed values listed in Table \ref{chemosensing parameters}.  The
parameters of the turn and roll modes are chosen randomly from
Gaussian distributions.  The average parameter values (indicated by
overbar) and standard deviations $\sigma$ are: normalized turning-mode
amplitude $\widebar{\amplitude}_2/\wavevector=1.67$ and
$\sigma_{\amplitude_2/\wavevector}=0.28$; turning-mode length
$\widebar{\Delta\wormcoord}\onp/\wormlength=0.5$ and
$\sigma_{\Delta\wormcoord\onp/\wormlength}=0.167$; normalized torsion
$\widebar{\torsion}_0\wormlength=2$ and
$\sigma_{\torsion_0\wormlength}=1$; and torsional-mode length
$\widebar{\Delta\wormcoord}\offp/\wormlength=0.5$ and
$\sigma_{\Delta\wormcoord\offp/\wormlength}=0.1$.  The parameter
distributions are the same for burrowing and swimming.

The turning rates listed in Table \ref{chemosensing parameters}
correspond to the directional persistence length of burrowing
trajectories $l_p\approx12\wormlength$ (which is of the same order as the
persistence length observed in crawling). We define
the persistence length $l_p$ as half-decay length of the directional
correlation function.  The rate of roll maneuvers was chosen such that
on average the worm spends half the time in a torsionless state.


{ \small
\paragraph*{\bf \small {Authors' contributions.}}
AB wrote the manuscript, developed locomotion and hydrodynamic codes,
and performed locomotion and chemotaxis simulations. AP provided
chemotaxis formulations and codes and performed some chemotaxis
simulations. MR provided experimental images of the worm.  SAV provided
experimental images and helped develop conceptual framework and draft
the manuscript.  JB conceived the study and wrote the manuscript.
\paragraph*{\bf \small Funding.}
AB would like to acknowledge financial support form NSF Grant CBET
1059745 and TTU Doctoral Dissertation
Completion Fellowship.  JB was partially supported from NSF Grant CBET
1603627.  SAV and MR acknowledge partial support from Grants NIH
1R21AG050503-01 and NASA NNX15AL16G.
\paragraph*{\bf \small Acknowledgments.}
We would like to thank our visiting undergraduate researcher, Jose
Montoya, for his help in developing chemotaxis codes and acknowledge
Dr.\ Szewczyk and Dr.\ Driscoll for useful discussions.
\paragraph*{\bf \small Competing interests.} 
We have no competing interests.
}

\clearpf
\bibliography{elegans}
\end{document}